\RequirePackage{fix-cm}
\documentclass[smallextended]{svjour3fix}       
\smartqed  
\usepackage{graphicx}
\usepackage{natbib}
\bibpunct[:]{(}{)}{;}{a}{}{,}

\usepackage{bm}
\usepackage{amsmath, amssymb}
\usepackage{bbm}
\usepackage{mathtools}
\usepackage{booktabs}
\usepackage{algorithm}
\usepackage{subcaption}
\usepackage{here}
\usepackage{rotating}
\usepackage{adjustbox}
\usepackage{empheq}
\usepackage{multirow}
\usepackage{tabularx}
\usepackage{arydshln}
\usepackage{url}
\def\eqref#1{(\ref{#1})}

\usepackage{mathtools} 
\mathtoolsset{
  showonlyrefs,
  showmanualtags,
}

\newcommand{\argmax}{\mathop{\rm arg~max}\limits}

\allowdisplaybreaks

\begin{document}

\title{
Misspecifying Non-Compensatory as Compensatory IRT: Analysis of Estimated Skills and Variance
}


\author{
    Hiroshi Tamano$^1$ \and
    Hideitsu Hino$^{2}$ \and
    Daichi Mochihashi$^2$ 
}


\institute{
    Hiroshi Tamano \at
   \email{hrs.tamano@gmail.com} 
    \and
    Hideitsu Hino \at
    \email{hino@ism.ac.jp}
    \and
    Daichi Mochihashi \at
    \email{daichi@ism.ac.jp}
    \and
    $^1$The Graduate University for Advanced Studies, SOKENDAI\\
    $^2$The Institute of Statistical Mathematics\\
}

\date{\empty}

\maketitle

\begin{abstract}
Multidimensional item response theory is a statistical test theory used to estimate the latent skills of learners and the difficulty levels of problems based on test results. Both compensatory and non-compensatory models have been proposed in the literature. Previous studies have revealed the substantial underestimation of higher skills when the non-compensatory model is misspecified as the compensatory model. However, the underlying mechanism behind this phenomenon has not been fully elucidated. It remains unclear whether overestimation also occurs and whether issues arise regarding the variance of the estimated parameters. In this paper, we aim to provide a comprehensive understanding of both underestimation and overestimation through a theoretical approach. In addition to the previously identified underestimation of the skills, we newly discover that the overestimation of skills occurs around the origin. Furthermore, we investigate the extent to which the asymptotic variance of the estimated parameters differs when considering model misspecification compared to when it is not taken into account.

\end{abstract}

\section{Introduction}
\label{introduction}
Item response theory (IRT) \citep{lord1952theory,rasch1960studies} is a statistical test theory used to estimate the skill level of learners and the difficulty of problems from test results. It can be used to evaluate learners consistently using the same criteria and is also employed in tests such as TOEIC and TOEFL. Multidimensional IRT (MIRT) \citep{reckase2009multidimensional} is an extension to deal with multiple skills and is divided into compensatory models \citep{reckase1985difficulty, reckase2009multidimensional} and non-compensatory models \citep{sympson1978model, whitely1980multicomponent, embretson2013multicomponent} based on the relationship between those skills. The compensatory models assume that multiple skills can compensate for each other, and the sum of the skills surpassing a certain threshold leads to a correct response. On the other hand, the non-compensatory models assume that multiple skills cannot compensate for each other, and each skill must exceed a certain threshold to yield a correct response.

The assumption of non-compensatory models is convincing in many situations. For example, understanding or solving an equation such as $x/5 + 3/10 = 2x$ requires both skills of fraction and equation; learners who only have either one cannot solve this equation. In this way, assigning skill tags to a problem with the ``and'' condition aligns with the non-compensatory assumption, whereas the ``or'' condition aligns with the compensatory assumption. Open datasets \citep{feng2009addressing, kddcup2010} provide sequential logs of item responses from learners, where problems are tagged with multiple skills. In these datasets, numerous problems are tagged using the ``and'' condition. Assigning skill tags to problems requires domain knowledge and is a time-consuming task; therefore, the estimation of appropriate skill tags from data has also been investigated \citep{oka2023scalable}.

Previous studies have demonstrated that fitting compensatory and non-compensatory models to data generated from non-compensatory models produces highly similar results \citep{spray1990comparison, christine2016partially}. \citet{christine2016partially} fitted both types of models to data generated from the non-compensatory model, assessed the errors between the estimated skill and the true skill, and revealed that the errors of these models were remarkably similar. This trend was particularly pronounced when the correlation between skills was high. Additionally, it was observed that the average error in the item response function was quite small even when fitted with the compensatory model.

While the previous studies have primarily examined average errors, \citet{buchholz2018impact} specifically focused on a subgroup of learners characterized by high proficiency in one skill and low proficiency in another skill. Their findings revealed a substantial underestimation of the higher skill, emphasizing the importance of employing the correct model, as learners in this subgroup can face significant disadvantages. Although their experimental evidence demonstrated this underestimation, the underlying mechanism has not been fully elucidated. It remains unclear whether underestimation occurs beyond the areas indicated in the previous study and if overestimation occurs as well. A theoretical approach is necessary to comprehensively understand the difference between the estimated skills and the true skills.

Furthermore, the variance of the estimated parameters is another concern for data analysts. When a model is correctly specified, the maximum likelihood estimator is asymptotically distributed with the variance being the inverse of Fisher information for a regular model. Since the latent skills are marginalized, the asymptotic normality is considered to be held in MIRT models. However, the asymptotic variance follows a different formula under the model misspecification \citep{white1982maximum}. If the difference between these variances is significant, data analysts need to be aware of model misspecification in evaluating the variance of the estimated parameters.

In this paper, we have two goals regarding this model misspecification. The first goal is to provide a comprehensive understanding of the underestimation and overestimation of skills using a theoretical approach. We approximate the direction from the true skills to the estimated skills using the gradient of the objective function at the true skills. By interpreting the gradient, we clarify the mechanism through which the difference between the estimated skills and the true skills arises. In addition to the underestimation previously identified in the literature, we newly discover that overestimation of skills occurs around the origin based on this mechanism. The second goal is to investigate the extent to which the asymptotic variance differs when model misspecification is considered versus when it is not considered. We derive the asymptotic variance by applying the result of \citet{white1982maximum} to our misspecified case. Simulation studies demonstrate that the asymptotic variances of both cases are quite close. Therefore, it turns out that underestimation or overestimation of the variance cannot be a critical issue in the assumed misspecified situation.

\section{Multidimensional Item Response Theory}\label{sec:mirt}

MIRT is a multi-skill extension of IRT that is used to estimate the skill level of learners and the difficulty of problems. We review two types of MIRT models: compensatory and non-compensatory models.

The input data of MIRT is a test result; it is typically expressed as binary matrix $Y$ whose $(i,j)$-th entry $y_{i,j} \in \{0,1\}$ represents the correct or incorrect answer to problem $i$ of learner $j$. The problems in the test are assumed to require multiple skills.

\begin{figure}
 \centering
 \begin{tabular}{cc}
  \begin{minipage}{0.4\linewidth}
  \includegraphics[width=\linewidth]{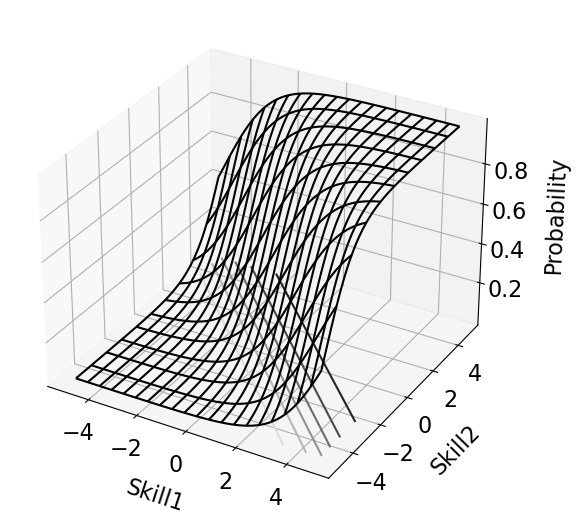}
  \end{minipage}
  &
  \begin{minipage}{0.4\linewidth}
  \includegraphics[width=\linewidth]{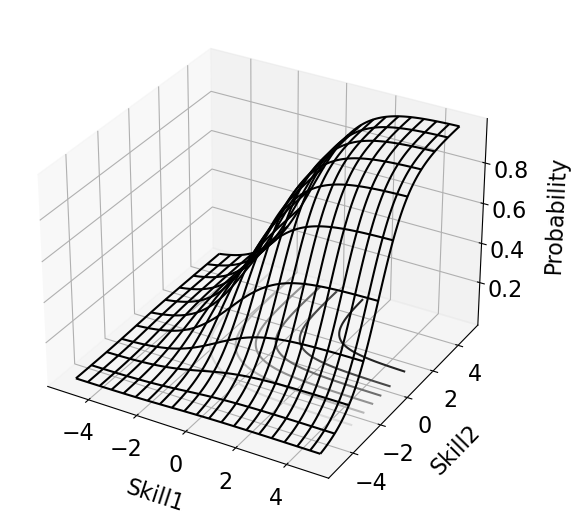}
  \end{minipage}\\
  (a) & (b) \\
  \end{tabular}
\caption{Comparison of item response surfaces between a compensatory
model and a non-compensatory model. (a) represents the compensatory
model and (b) represents the non-compensatory model. The probability
is high when either skill is high in (a), whereas it is high
when both skills are high in (b).
}
\label{pic:cm_ncm_diff}
\end{figure}

MIRT involves compensatory models and non-compensatory models. Compensatory models assume that each skill can complement other skills. In the compensatory model, the probability that learner $j$ can solve problem $i$ is defined as
\begin{equation}\label{eqn:cm_p_y1}
    p_c(y_{i,j}=1|\bm{\gamma_j}, \bm{\alpha_i}, \beta_i) = 
    \frac{1}{1+\exp(-(\sum_{k=1}^K \alpha_{i,k}\gamma_{j,k} + \beta_i))},
\end{equation}
where $\bm{\gamma}_j$ is a $K$-dimensional vector of latent skills for learner $j$, $\bm{\alpha}_i$ is a $K$-dimensional vector of the discrimination parameters of problem $i$, and $\beta_i$ is the difficulty parameter of problem $i$. $K$ denotes the number of skills. This is called the two-parameter logistic model because of the discrimination and difficulty parameters. The latent skill is summed up; thus, each skill could complement other skills. Fig.~\ref{pic:cm_ncm_diff} (a) shows a two-dimensional example of Eq.~\eqref{eqn:cm_p_y1}.

In contrast, non-compensatory models assume that individual skills cannot complement other skills. In the non-compensatory model, the probability that learner $j$ can solve problem $i$ is defined as
\begin{equation}\label{eqn:ncm_p_y1}
    p_n(y_{i,j}=1|\bm{z_j}, \bm{a_i}, \bm{b_i}) = 
    \prod_{k=1}^{K} \frac{1}{1+\exp(-a_{i,k} (z_{j,k} - b_{i,k}))},
\end{equation}
where $\bm{z}_j$ is a $K$-dimensional vector of latent skills for learner $j$, $\bm{a}_i$ is a $K$-dimensional vector of the discrimination parameters of problem $i$, and $\bm{b}_i$ is a $K$-dimensional vector of the difficulty parameters of problem $i$. Here, the probability is multiplied; thus, individual skills cannot complement other skills. Fig.~\ref{pic:cm_ncm_diff} (b) shows a two-dimensional example of Eq.~\eqref{eqn:ncm_p_y1}.

\section{Setting of Model Misspecification}\label{sec:misspecification}

Let us assume that the true data-generating model is the non-compensatory model defined in Eq.~\eqref{eqn:ncm_p_y1}. When the number of problems is $M$, a response variable $\bm{y} \in \{0, 1\}^M$ for a leaner is drawn from
\begin{align}
  p_n(\bm{y}| \bm{a}, \bm{b})
  &= \int \prod_{i=1}^M 
  p_n(y_i| \bm{a}_i, \bm{b}_i, \bm{z}) \mathcal{N}(\bm{z}|0, \rm{I}) d\bm{z},
  \label{eqn:ncm_p_y}
\end{align}
where the latent skills of the learner are assumed to be distributed in the standard normal distribution and are marginalized out. Here, $\bm{a}$ represents $\{\bm{a}_1, ..., \bm{a}_M\}$ and $\bm{b}$ represents $\{\bm{b}_1, ..., \bm{b}_M\}$.

On the other hand, the model used for fitting to data is the compensatory model defined in Eq.~\eqref{eqn:cm_p_y1}. By including both cases of $y_i=0$ and $y_i=1$, Eq.~\eqref{eqn:cm_p_y1} can be rewritten as
\begin{align}
  p_c(y_i| \bm{\alpha}_i, \beta_i, \bm{\gamma})
  & = \frac{\exp\{y_i (\bm{\alpha}_i^{\top} \bm{\gamma} + \beta_i )\}}
           {1 + \exp{\{\bm{\alpha}_i^{\top} \bm{\gamma} + \beta_i\}}}.
\end{align}
As seen in the non-compensatory model, marginalizing the latent skills by the standard normal distribution gives
\mathtoolsset{showonlyrefs=false}
\begin{equation}
  p_c(\bm{y}|\bm{\alpha}, \bm{\beta})
  = \int \prod_{i=1}^{M}
    p_c(y_i| \bm{\alpha}_i, \beta_i, \bm{\gamma}) \mathcal{N}(\bm{\gamma}|\bm{0}, \rm{I})
    d\bm{\gamma}
  \approx
  \sum_{l} w(\bm{\gamma}_l) \prod_{i=1}^{M}
  p_c(y_i| \bm{\alpha}_i, \beta_i, \bm{\gamma}_l),
  \label{eqn:cm_p_y_quadrature}
\end{equation}
\mathtoolsset{showonlyrefs=true}
where $\bm{\alpha}$ represents $\{\bm{\alpha}_1, ..., \bm{\alpha}_M\}$ and $\bm{\beta}$ represents $\{\beta_1, ..., \beta_M\}$. This marginalization is approximated by the Gauss-Hermite quadrature \citep{jackel2005note}; $\bm{\gamma}_l$ denotes a quadrature point and $w(\bm{\gamma}_l)$ denotes a weight for the point. When the responses for $N$ learners are given, our objective function of fitting the compensatory model is given as $p_c(Y| \bm{\alpha}, \bm{\beta})=\prod_{j=1}^{N} p_c(\bm{y}_j|\bm{\alpha}, \bm{\beta}),$ where $\bm{y}_j$ denotes the response vector of learner~$j$. This can be maximized with respect to $\bm{\alpha}$ and $\bm{\beta}$ using the Expectation-Maximization (EM) algorithm \citep{bock1981marginal}.

\section{Difference between Estimated Skills and True Skills}\label{sec:difference}

We elucidate the mechanism behind the occurrence of the difference between the estimated skills and the true skills in the misspecified situation. Given the item parameters maximizing the marginal likelihood, we assume that skills are estimated using the maximum a posteriori (MAP) estimation. Initially, we approximate the direction from the true skills to the estimated skills with a gradient. By interpreting the gradient, we describe the mechanism through which skills are either underestimated or overestimated.

\subsection{Approximating Difference with Gradient}

Here, we derive the gradient of the objective function of the compensatory model at the true skills to approximate the direction from the true skills to the estimated skills.

The log-likelihood of responses with a prior for a learner in the compensatory model is given by
\begin{align}
  L_c(\bm{\gamma})
  & = \sum_{i=1}^M \ln p_c(y_i| \bm{\alpha}_i, \beta_i, \bm{\gamma}) 
  + \ln \mathcal{N}(\bm{\gamma}| \bm{0}, \rm{I})\\
  & = \sum_{i=1}^M \left\{
  y_i (\bm{\alpha}_i^{\top} \bm{\gamma} + \beta_i )
  - \ln{
    \left(
      1 + \exp{(\bm{\alpha}_i^{\top} \bm{\gamma} + \beta_i)}
    \right)
  }
  \right\}
  + \ln \mathcal{N}(\bm{\gamma}| \bm{0}, \rm{I}).
  \label{eqn:likelihood_learner_cm}
\end{align}
The gradient with respect to the skill parameters $\bm{\gamma}$ is calculated as
\begin{align}
  \nabla_{\bm{\gamma}} L_c(\bm{\gamma})
  & = \sum_{i=1}^M \left\{
  y_i 
  - \frac{\exp{(\bm{\alpha}_i^{\top} \bm{\gamma} + \beta_i)}}
         {1 + \exp{(\bm{\alpha}_i^{\top} \bm{\gamma} + \beta_i)}}
  \right\}
  \bm{\alpha}_i - \bm{\gamma}.
\end{align}
By plugging $\bm{\alpha}^*$ and $\bm{\beta}^*$ into $\nabla_{\bm{\gamma}} L_c(\bm{\gamma})$, we have
\begin{align}
  \nabla_{\bm{\gamma}} L_c^*(\bm{\gamma})
  & = \sum_{i=1}^M \left\{
  y_i 
  - \frac{\exp{(\bm{\alpha}_i^{*\top} \bm{\gamma} + \beta_i^*)}}
         {1 + \exp{(\bm{\alpha}_i^{*\top} \bm{\gamma} + \beta_i^*)}}
  \right\}
  \bm{\alpha}_i^* - \bm{\gamma},
  \label{eqn:gradient_learner_cm}
\end{align}
where $\bm{\alpha}^*$ and $\bm{\beta}^*$ are the values obtained by fitting the compensatory model to the data as
$
  \bm{\alpha}^*, \bm{\beta}^*
  =
  \argmax \int p_n(\bm{y}|\bm{a}^*, \bm{b}^*)
  \ln p_c(\bm{y}|\bm{\alpha}, \bm{\beta}) d\bm{y}.
$
$\bm{a}^*$ and $\bm{b}^*$ are the true model parameters of the non-compensatory model that generates the data.
Eq.~\eqref{eqn:gradient_learner_cm} approximately represents the direction from any $\bm{\gamma}$ to $\bm{\gamma}^*$ and $\bm{\gamma}^*$ is the MAP estimation in the compensatory model. Because $\bm{\alpha}^*$ and $\bm{\beta}^*$ are not available in practice, we use
$
  \hat{\bm{\alpha}}, \hat{\bm{\beta}}
  =
  \argmax \sum_{j=1}^N
  \ln p_c(\bm{y}_j|\bm{\alpha}, \bm{\beta})
$
with large number of $N$ samples $\{\bm{y}_1, ..., \bm{y}_N\}$ from $p_n(\bm{y}|\bm{a}^*, \bm{b}^*)$ to approximate $\bm{\alpha}^*, \bm{\beta}^*$.

\begin{figure}[tb]
  \centering
  \includegraphics[width=0.3\linewidth]{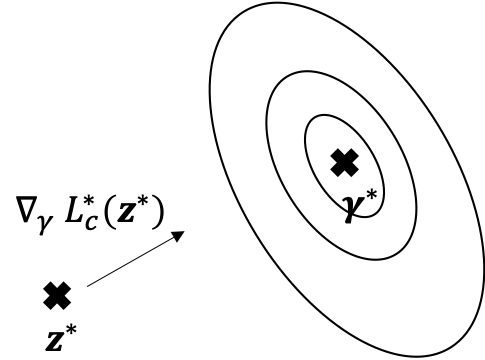}
\caption{The gradient at the point of the true skills with respect to the objective function of the compensatory model approximates how the skills are differently estimated against the true skills. The true skills are denoted as $\bm{z}^*$ and the estimated skills are denoted as $\bm{\gamma}^*$.}
\label{pic:gradient_z_star}
\end{figure}

Here, we consider $\nabla_{\bm{\gamma}} L_c^*(\bm{z}^*)$ which is the gradient at $\bm{\gamma} = \bm{z}^*$. Because $\bm{z}^*$ is the true skills in the non-compensatory model, $\nabla_{\bm{\gamma}} L_c^*(\bm{z}^*)$ approximately represents the direction from $\bm{z}^*$ to $\bm{\gamma}^*$ as shown in Fig.~\ref{pic:gradient_z_star}. For any given skills $\bm{z}$ in the non-compensatory model, the expected direction from the true skill to the estimated skills can be calculated as
\begin{align}
  \mathbb{E}_{\bm{y}|\bm{z}}[\nabla_{\bm{\gamma}} L^*_c(\bm{z})]
  & = \sum_{i=1}^M \left\{
  \prod_{k=1}^K
  \left(
    \frac{1}{1 + \exp{\{-a_{i,k}^* (z_k - b_{i,k}^*)\}}}
  \right)
  \right. \nonumber \\
  & \quad
  \left.
  - \frac{\exp{(\bm{\alpha}_i^{*\top} \bm{z} + \beta_i^*)}}
         {1 + \exp{(\bm{\alpha}_i^{*\top} \bm{z} + \beta_i^*)}}
  \right\}
  \bm{\alpha}_i^* - \bm{z} \nonumber \\
  & =
  \sum_{i=1}^M \left\{
    p_n(y_i=1|\bm{z}, \bm{a}_i^*, \bm{b}_i^*)
    - p_c(y_i=1|\bm{z}, \bm{\alpha}_i^*, \beta_i^*) 
  \right\} \bm{\alpha}_i^* - \bm{z}
  \label{eqn:expected_gradient_learner_cm}
\end{align}


\subsection{Interpreting Gradient}

\begin{figure}
 \centering
  \includegraphics[width=0.4\linewidth]{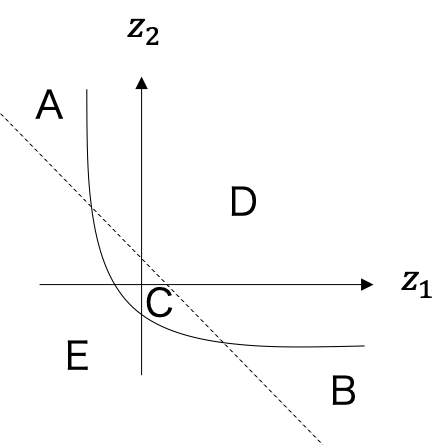}
  \caption{Five areas A-E to explain the underestimation and overestimation effect. The curved line denotes the classification boundary in the non-compensatory model, whereas the straight dashed line denotes the boundary in the compensatory model. Mainly, areas A and B take the underestimation effect and area C takes the overestimation effect.}
  \label{pic:five_area}
\end{figure}

Now that Eq.~\eqref{eqn:expected_gradient_learner_cm}, which approximates the direction from the true skills to the estimated skills, is derived, we interpret it and discuss the mechanism of underestimation and overestimation of skills. Because the last term $-\bm{z}$ in Eq.~\eqref{eqn:expected_gradient_learner_cm} comes from a prior distribution, we interpret Eq.~\eqref{eqn:expected_gradient_learner_cm} except $-\bm{z}$. Eq.~\eqref{eqn:expected_gradient_learner_cm} is the sum over problems; we look into the effect caused by each problem. Since discrimination parameters $\bm{\alpha}_i^*$ are positive, the underestimation or overestimation is determined by the sign of
\[
  p_n(y_i=1|\bm{z}, \bm{a}_i^*, \bm{b}_i^*)
  - p_c(y_i=1|\bm{z}, \bm{\alpha}_i^*, \beta_i^*) \quad (:= p_n - p_c).
\]
The first term is the probability of $y_i=1$ in the true data-generating model and the second term is the one in the fitted compensatory model. We abbreviate the first term as $p_n$, and the second term as  $p_c$. Fig.~\ref{pic:five_area} shows the contours of probability $0.5$ of item response surfaces for compensatory and non-compensatory models. In areas A and B, $p_n$ is smaller than $0.5$ and $p_c$ is larger than $0.5$; the two models predict the response differently. Since $p_n - p_c$ is negative, learners around these areas take an underestimation effect from this problem. In area C, $p_n$ is larger than $0.5$ and $p_c$ is smaller than $0.5$. Since $p_n - p_c$ is positive, learners around this area take an overestimation effect. However, area C is considered to be not so wide. Because both compensatory and non-compensatory models assume the standard normal distribution on the skill parameters, the scale of skill is the same. The classification boundary tends to be placed around the same. In areas D and E, both $p_n$ and $p_c$ are over $0.5$ or under $0.5$ and the difference is small. The effect in these areas is considered to be ignorable. By interpreting Eq.~\eqref{eqn:expected_gradient_learner_cm}, the difference between the estimated skills and the true skills is considered to come mainly from areas A, B, and C.

\begin{figure}
 \centering
 {\tabcolsep=0pt
 \begin{tabular}{cc}
  \begin{minipage}{0.4\linewidth}
  \includegraphics[width=\linewidth]{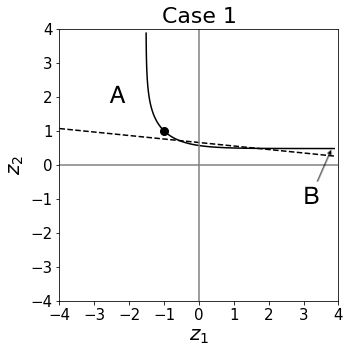}
  \end{minipage}
  &
  \begin{minipage}{0.4\linewidth}
  \includegraphics[width=\linewidth]{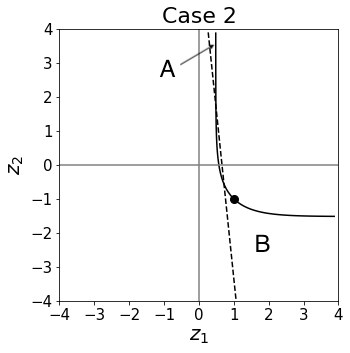}
  \end{minipage}\\
  \begin{minipage}{0.4\linewidth}
  \includegraphics[width=\linewidth]{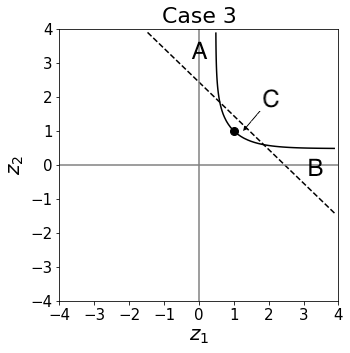}
  \end{minipage}
  &
  \begin{minipage}{0.4\linewidth}
  \includegraphics[width=\linewidth]{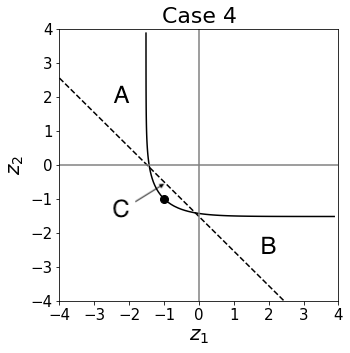}
  \end{minipage}\\
  \end{tabular}
  }
\caption{Four cases of items to explain the effect for skill estimation. The black dots denotes $[z_1^{(0.5)}, z_2^{(0.5)}]^{\top}$. Case 1 mainly has the underestimation effect for skill 2 in area A. Case 2 mainly has the underestimation effect for skill 1 in area B. Cases 3 and 4 have the equal underestimation effect for skills 1 and 2 in areas A and B; area C has the overestimation effect for both skills.}
\label{pic:ncm_item_four_cases}
\end{figure}

Finally, we categorize items in the non-compensatory model into four cases and examine areas A, B, and C for each case in detail because those areas differ depending on each case. Fig.~\ref{pic:ncm_item_four_cases} illustrates the four cases of items. For an item with discrimination parameters $[a_1, a_2]^{\top}$ and difficulty parameters $[b_1, b_2]^{\top}$, the skills $[z_1^{(0.5)}, z_2^{(0.5)}]^{\top}$ for which the correct probability is 0.5 can be calculated as:
$
   z_1^{(0.5)} = b_1 + \frac{1}{a_1}\sigma^{-1}(\sqrt{0.5})
$ and
$
   z_2^{(0.5)} = b_2 + \frac{1}{a_2}\sigma^{-1}(\sqrt{0.5}).
$

In Case 1, $z_1^{(0.5)} < 0$ and $z_2^{(0.5)} > 0$. Considering that the skill is assumed to follow a standard normal distribution, the classification boundary in the compensatory model tends to approximate the boundary in the non-compensatory model by weighting around the origin. Consequently, the discrimination parameters in the compensatory model should satisfy $\alpha_1 < \alpha_2$ to indicate that area A dominates over area B. Referring back to Eq.~\eqref{eqn:expected_gradient_learner_cm}, the value for each skill is weighted by the discrimination parameter of the compensatory model. When $\alpha_1 < \alpha_2$, it implies that skill 2 is primarily affected by the underestimation effect from this item. The main effect of items categorized in Case 1 is the underestimation of skill 2 for learners around area A.

In Case 2, $z_1^{(0.5)} > 0$ and $z_2^{(0.5)} < 0$. Since the classification boundary in the non-compensatory model tends to be locally approximated by the compensatory model around the origin, we have $\alpha_2 < \alpha_1$. Consequently, area B dominates over area A, and the underestimation effect mainly affects skill 1. The main effect of the item categorized in Case 2 is the underestimation of skill 1 for learners around area B.

In Case 3, $z_1^{(0.5)} > 0$ and $z_2^{(0.5)} > 0$. In this case, the weights assigned to $\alpha_1$ and $\alpha_2$ tend to be equal. There is an underestimation effect for both skills around areas A and B, while an overestimation effect for both skills appears around area C.

In Case 4, $z_1^{(0.5)} < 0$ and $z_2^{(0.5)} < 0$. Similar to Case 3, the weights assigned to $\alpha_1$ and $\alpha_2$ tend to be equal. There is an underestimation effect for both skills around areas A and B, while an overestimation effect for both skills emerges around area C.

By considering various items within each case, learners with $z_1 < 0$ and $z_2 > 0$ tend to experience the underestimation effect for skills 1 and 2, with skill 2 being more affected by the items categorized in Case 1. Learners with $z_1 > 0$ and $z_2 < 0$ encounter the underestimation effect for skills 1 and 2, with skill 1 being more affected by the items categorized in Case 2. On the other hand, learners around the origin tend to experience a slight overestimation effect for both skills.

\section{Asymptotic Variance of Estimated Parameters}\label{sec:variance}

In this section, we derive the asymptotic variances of the item parameters estimated by the maximum likelihood estimation (MLE) under the assumed model misspecification. We focus on the item parameters because learners' skills are typically marginalized out with the standard normal distribution and item parameters are estimated in MLE. 

Based on the result in \cite{white1982maximum}, as $n \to \infty$, the following properties hold.
\begin{align}
  \sqrt{n}(\hat{\beta}_h^{(n)} - \beta_h^{*})
  & \stackrel{D}\to
  \mathcal{N}(0, I(\beta_h^{*})^{-1} J(\beta_h^{*}) I(\beta_h^{*})^{-1}), 
  \label{eqn:asymptotic_var_beta} \\
  \sqrt{n}(\hat{\alpha}_{h,s}^{(n)} - \alpha_{h,s}^{*})
  & \stackrel{D}\to
  \mathcal{N}(0, I(\alpha_{h,s}^{*})^{-1} J(\alpha_{h,s}^{*}) I(\alpha_{h,s}^{*})^{-1}).
  \label{eqn:asymptotic_var_alpha}
\end{align}
$\hat{\beta}_h^{(n)}$, $\beta_h^{*}$, $I(\beta_h^{*})$, $J(\beta_h^{*})$, $I(\alpha_{h,s}^{*})$, and $J(\alpha_{h,s}^{*})$ are define as
\begin{align}
  \hat{\beta}_h^{(n)}, \hat{\alpha}_{h,s}^{(n)} 
  & = \mbox{argmax} \sum_{j=1}^{n} \ln p_c(\bm{y}_j|\bm{\alpha}, \bm{\beta}), 
  \label{eqn:empirical_solution_in_cm} \\
  \hat{\beta}_h^{*}, \hat{\alpha}_{h,s}^{*} 
  & = \mbox{argmax} \int p_n(\bm{y}) \ln p_c(\bm{y}|\bm{\alpha}, \bm{\beta}) d\bm{y}, \\
  I(\beta_h^{*})
  &= - \mathbb{E}_{p_n(\bm{y})}\left[
    \left.
    \frac{\partial^2}{\partial \beta_h^2}
    \ln p_c(\bm{y}|\bm{\alpha}, \bm{\beta})
    \right|_{\beta_h = \beta_h^*}
  \right], \\
  J(\beta_h^{*})
  &= \mathbb{E}_{p_n(\bm{y})}\left[\left(
    \left.
    \frac{\partial}{\partial \beta_h} \ln p_c(\bm{y}|\bm{\alpha}, \bm{\beta})
    \right|_{\beta_h = \beta_h^*}
  \right)^2 \right], \\
  I(\alpha_{h,s}^{*})
  &= - \mathbb{E}_{p_n(\bm{y})}\left[
    \left.
    \frac{\partial^2}{\partial \alpha_{h,s}^2}
    \ln p_c(\bm{y}|\bm{\alpha}, \bm{\beta})
    \right|_{\alpha_{h,s} = \alpha_{h,s}^*}
  \right], \\
  J(\alpha_{h,s}^{*})
  &= \mathbb{E}_{p_n(\bm{y})}\left[\left(
    \left.
    \frac{\partial}{\partial \alpha_{h,s}} \ln p_c(\bm{y}|\bm{\alpha}, \bm{\beta})
    \right|_{\alpha_{h,s} = \alpha_{h,s}^*}
  \right)^2 \right].
\end{align}
The complete calculation of the first and second derivatives of $\ln p_c(\bm{y}|\bm{\alpha}, \bm{\beta})$ with respect to $\beta_h$ and $\alpha_{h,s}$ are provided in the supplemental material.

\section{Simulation Study of Difference}\label{sec:simulation_difference}

We conducted an experiment to assess the accuracy of $\nabla_{\bm{\gamma}} L_c^*(\bm{z}^*)$ in approximating the true direction of the difference $\bm{\gamma}^* - \bm{z}^*$. A true non-compensatory model was set by specifying $\{\bm{a}^*, \bm{b}^*, \bm{z}^*\}$ and $\bm{y}$ was generated by using Eq.~\eqref{eqn:ncm_p_y} with $K=2$. Then $\hat{\bm{\alpha}}$, $\hat{\bm{\beta}}$, and $\hat{\bm{\gamma}}$ were obtained by fitting the compensatory model to approximate $\bm{\alpha}^*$, $\bm{\beta}^*$, and $\bm{\gamma}^*$. We show the correlation between $\nabla_{\bm{\gamma}} L_c^*(\bm{z}^*)$ and the actual difference $\bm{\gamma}^* - \bm{z}^*$ and visualize the actual direction of the difference and the gradient. After seeing that $\nabla_{\bm{\gamma}} L_c^*(\bm{z}^*)$ approximately describes the actual direction, we visualize Eq.~\eqref{eqn:expected_gradient_learner_cm} to provide a comprehensive understanding of the entire difference.

\subsection{Data Generation}

\begin{figure}
 \centering
  \includegraphics[width=0.3\linewidth]{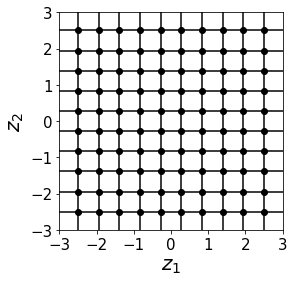}
\caption{100 problems requiring two skills. Each difficulty parameter is set to place $0.5$ probability to each point in the lattice, which corresponds to $[z_1^{(0.5)}, z_2^{(0.5)}]^{\top}$. This guarantees that the problems of cases 1-4 exist equally.}
\label{pic:lattice_b}
\end{figure}

We created a total of 120 problems for this experiment, with 10 problems for skill 1, 10 problems for skill 2, and 100 problems requiring both skill 1 and skill 2. The number of skills is 2. The item discrimination parameters $a_{i,k}$ were sampled from a log-normal distribution, specifically $\mbox{LogN}(0.2, 0.2)$. For the problems that only require one skill, the difficulty parameters ranged from $-2.5$ to $2.5$, evenly spaced. As for the problems requiring two skills, their difficulty parameters were placed on a lattice structure as depicted in Fig.~\ref{pic:lattice_b}. Each point on the lattice represents $[z_1^{(0.5)}, z_2^{(0.5)}]^{\top}$, which corresponds to the four cases (Cases 1-4) discussed in Section \ref{sec:difference}, and these cases are evenly distributed.

We generated 1,000,000 learners by drawing the skill parameters $\bm{z}_j$ from the standard normal distributions and drawing $\bm{y}_j$ from the Bernoulli distribution using Eq.~\eqref{eqn:ncm_p_y1}. The parameters in the compensatory model $\bm{\alpha}_i$, $\bm{\beta}_i$, and $\bm{\gamma}_j$ were obtained by fitting the model to $Y=[\bm{y}_1, \dots, \bm{y}_N]$ using {\tt mirt} package in R \citep{chalmers2012mirt}.
With all the necessary components in place, we are now equipped to calculate $\nabla{\bm{\gamma}} L_c^*(\bm{z}^*)$.

%

\subsection{Results}

\begin{figure}
 \centering
 {\tabcolsep=0pt
 \begin{tabular}{cc}
  \begin{minipage}{0.35\linewidth}
  \includegraphics[width=\linewidth]{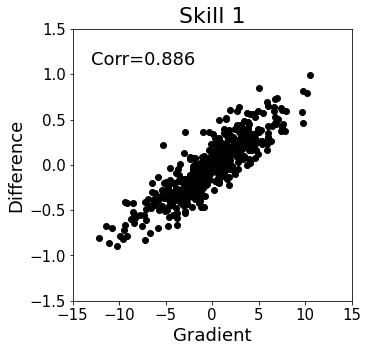}
  \end{minipage}
  &
  \begin{minipage}{0.35\linewidth}
  \includegraphics[width=\linewidth]{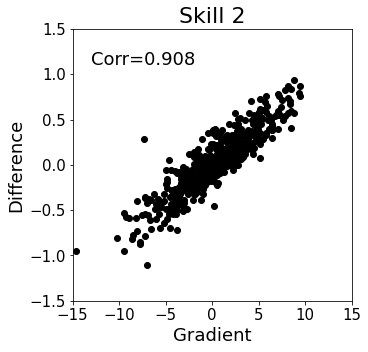}
  \end{minipage}
  \end{tabular}
  }
\caption{Scatter plot between the gradient (x-axis) and the difference between the estimated skills and the true skills (y-axis). The gradient and the difference are highly correlated.}
\label{pic:scatter_gradient_true_bias}
\end{figure}

First, we present a scatter plot in Fig.~\ref{pic:scatter_gradient_true_bias} that illustrates the relationship between the gradient $\nabla_{\bm{\gamma}} L_c^*(\bm{z}^*)$ and the difference $\bm{\gamma}^* - \bm{z}^*$, with the left plot corresponding to skill 1 and the right plot corresponding to skill 2. It is evident that the gradient and the actual difference exhibit a strong positive correlation, with Pearson's correlations approximately equal to $0.9$ for both skills.

\begin{figure}
 \centering
 {\tabcolsep=0pt
 \begin{tabular}{cc}
  \begin{minipage}{0.35\linewidth}
  \includegraphics[width=\linewidth]{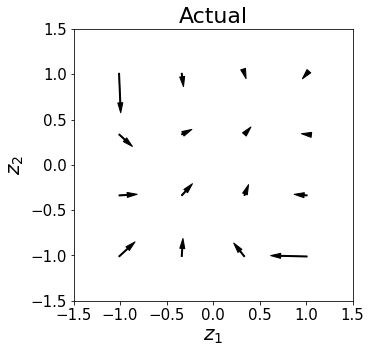}
  \end{minipage}
  &
  \begin{minipage}{0.35\linewidth}
  \includegraphics[width=\linewidth]{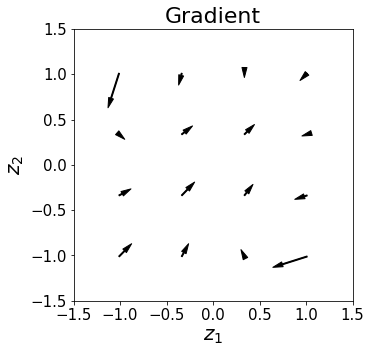}
  \end{minipage}
  \end{tabular}
  }
\caption{The actual directions from the true skills to the estimated skills (left) and the gradient (right) for $4 \times 4$ areas with approximately equal samples. The strongest underestimation effect is shown in the left upper corner and the right lower corner for both the actual and the gradient. The actual only has the underestimation effect for skill 2 in the left upper corner and the underestimation effect for skill 1 in the right lower corner.}
\label{pic:bias_arrow}
\end{figure}

Next, we analyze the direction of the difference $\bm{\gamma}^* - \bm{z}^*$ and the gradient by dividing the skill space into 16 regions, each containing a roughly equal number of samples. Fig.~\ref{pic:bias_arrow} presents the average direction of the difference (left) and the gradient (right) for each region. Notably, both the actual and the gradient display a pronounced underestimation effect in the upper left and lower right corners. However, there is a difference between the two: in the upper left corner, the actual underestimates only skill 2, whereas the gradient underestimates both skill 1 and skill 2. Similarly, in the lower right corner, the actual underestimates only skill 1, while the gradient underestimates both skills. Although the gradient underestimates both skills, the magnitudes differ. Tables~\ref{tab:gradient_skill_1} and \ref{tab:gradient_skill_2} provide the values of the gradient for skills 1 and 2, respectively. In the upper left corner, the gradient for skill 2 is $-5.176$, whereas for skill 1 it is $-1.680$. In the lower right corner, the gradient for skill 1 is $-5.209$, while for skill 2 it is $-1.629$. The discussion in Section \ref{sec:difference} provides an explanation for this behavior, where items of case 1 exert a stronger underestimation effect on skill 2, and items of case 2 exhibit a stronger underestimation effect on skill 1. Apart from the upper left and lower right corners, the actual and the gradient align closely. We observe an overestimation effect for both skills along the diagonal line starting from the lower left corner. Additionally, Tables~\ref{tab:true_bias_skill_1} and \ref{tab:true_bias_skill_2} present the differences between the estimated skills and the true skills for skills 1 and 2, respectively, complementing the information in Fig.~\ref{pic:bias_arrow}.

\begin{figure}
 \centering
 {\tabcolsep=0pt
 \begin{tabular}{cc}
  \begin{minipage}{0.35\linewidth}
  \includegraphics[width=\linewidth]{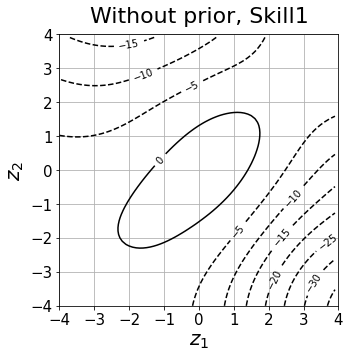}
  \end{minipage}
  &
  \begin{minipage}{0.35\linewidth}
  \includegraphics[width=\linewidth]{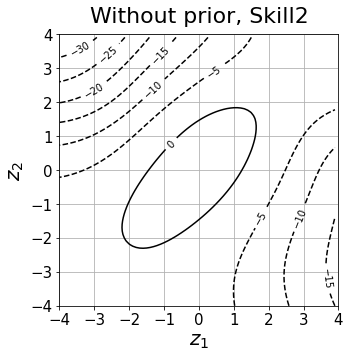}
  \end{minipage}
  \\
  \begin{minipage}{0.35\linewidth}
  \includegraphics[width=\linewidth]{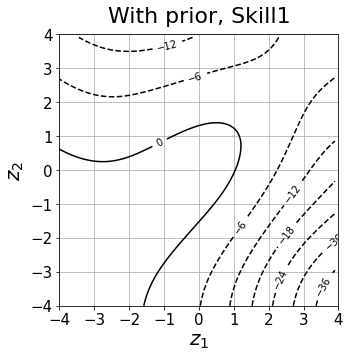}
  \end{minipage}
  &
  \begin{minipage}{0.35\linewidth}
  \includegraphics[width=\linewidth]{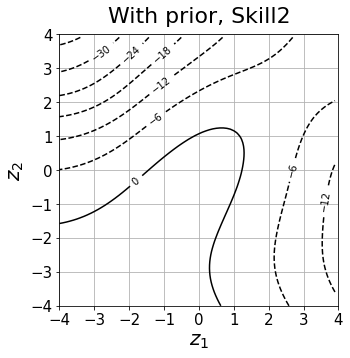}
  \end{minipage}
  
  \end{tabular}
  }
\caption{The expected gradient in Eq.~\eqref{eqn:expected_gradient_learner_cm} with and without prior. Skill 1 is strongly underestimated in the right lower corner and weakly underestimated in the left upper corner for both cases with and without prior. Skill 2 is strongly underestimated in the left upper corner and weakly underestimated in the right lower corner for both cases with and without prior. A slight overestimation for both skills can be seen around the origin when the prior is not considered and it is shifted towards the left lower corner when the prior is considered.}
\label{pic:expected_gradient}
\end{figure}

Finally, we visualize Eq.~\eqref{eqn:expected_gradient_learner_cm} with and without considering the prior in Fig.~\ref{pic:expected_gradient}. 
Eq.~\eqref{eqn:expected_gradient_learner_cm} explains that skill 1 is strongly underestimated in the right lower corner and weakly underestimated in the left upper corner for both cases with and without prior. The former coincides with the actual direction of the difference; however, the latter is not observed in the actual. This can be considered the limitation of the first derivative. Skill 2 is strongly underestimated in the left upper corner and weakly underestimated in the right lower corner for both cases with and without prior. Likewise skill 1, the latter is not observed in the actual.
When the prior is not considered, a slight overestimation effect is visible for both skills around the origin, which shifts towards the lower left corner when the prior is taken into account. This overestimation aligns with the actual observations.

\begin{table}[tb]
    \centering
    \caption{Difference between the estimated skills and the true skills in skill 1 for each $4 \times 4$ area. The strong underestimation can be seen in the right lower corner.}
    \label{tab:true_bias_skill_1}
\begin{tabular}{lrrrr}
\toprule
& \multicolumn{4}{c}{Skill 1} \\
\cmidrule(r){2-5}
Skill 2 & (-inf, -0.674] & (-0.674, 0.0] & (0.0, 0.674] & (0.674, inf] \\
\midrule
(0.674, inf] & 0.018 & 0.019 & 0.018 & -0.051 \\
(0.0, 0.674] & 0.144 & 0.109 & 0.071 & -0.061 \\
(-0.674, 0.0] & 0.195 & 0.115 & 0.045 & -0.139 \\
(-inf, -0.674] & 0.172 & 0.013 & -0.115 & -0.392 \\
\bottomrule
\end{tabular}
\end{table}

\begin{table}[tb]
    \centering
    \caption{Difference between the estimated skills and the true skills in skill 2 for each $4 \times 4$ area. The strong underestimation can be seen in the left upper corner.}
    \label{tab:true_bias_skill_2}
\begin{tabular}{lrrrr}
\toprule
& \multicolumn{4}{c}{Skill 1} \\
\cmidrule(r){2-5}
Skill 2 & (-inf, -0.674] & (-0.674, 0.0] & (0.0, 0.674] & (0.674, inf] \\
\midrule
(0.674, inf] & -0.436 & -0.147 & -0.059 & -0.060 \\
(0.0, 0.674] & -0.134 & 0.055 & 0.082 & 0.007 \\
(-0.674, 0.0] & 0.013 & 0.130 & 0.119 & 0.012 \\
(-inf, -0.674] & 0.161 & 0.199 & 0.148 & 0.010 \\
\bottomrule
\end{tabular}
\end{table}

\begin{table}[tb]
    \centering
    \caption{Gradient in skill 1 for each $4 \times 4$ area. The strong underestimation effect can be seen in the right lower corner.}
    \label{tab:gradient_skill_1}
\begin{tabular}{lrrrr}
\toprule
& \multicolumn{4}{c}{Skill 1} \\
\cmidrule(r){2-5}
Skill 2 & (-inf, -0.674] & (-0.674, 0.0] & (0.0, 0.674] & (0.674, inf] \\
\midrule
(0.674, inf] & -1.680 & -0.472 & -0.007 & -1.079 \\
(0.0, 0.674] & 0.865 & 1.641 & 1.527 & -0.783 \\
(-0.674, 0.0] & 1.813 & 1.960 & 1.303 & -1.876 \\
(-inf, -0.674] & 1.916 & 1.046 & -0.523 & -5.209 \\
\bottomrule
\end{tabular}
\end{table}

\begin{table}[tb]
    \centering
    \caption{Gradient in skill 2 for each $4 \times 4$ area. The strong underestimation effect can be seen in the left upper corner.}
    \label{tab:gradient_skill_2}
\begin{tabular}{lrrrr}
\toprule
& \multicolumn{4}{c}{Skill 1} \\
\cmidrule(r){2-5}
Skill 2 & (-inf, -0.674] & (-0.674, 0.0] & (0.0, 0.674] & (0.674, inf] \\
\midrule
(0.674, inf] & -5.176 & -1.765 & -0.630 & -1.082 \\
(0.0, 0.674] & -0.743 & 1.269 & 1.479 & -0.242 \\
(-0.674, 0.0] & 0.983 & 2.007 & 1.658 & -0.588 \\
(-inf, -0.674] & 1.923 & 1.921 & 1.032 & -1.629 \\
\bottomrule
\end{tabular}    
\end{table}

\section{Simulation Study of Asymptotic Variance}\label{sec:simulation_variance}

We compute the experimental asymptotic variances (the left-hand sides in Eq.~\eqref{eqn:asymptotic_var_beta} and Eq.~\eqref{eqn:asymptotic_var_alpha}), the theoretical asymptotic variances considering misspecification (the right-hand side in Eq.~\eqref{eqn:asymptotic_var_beta} and Eq.~\eqref{eqn:asymptotic_var_alpha}) denoted as $I^{-1}JI^{-1}$, and the theoretical asymptotic variances without considering misspecification denoted as $I^{-1}$. When data analysts do not notice the model misspecification, they use $I^{-1}$ as the asymptotic variance because it is correct under the correctly specified model.
Firstly, we compare the experimental asymptotic variance with $I^{-1}JI^{-1}$ to assess the validity of Eq.~\eqref{eqn:asymptotic_var_beta} and Eq.~\eqref{eqn:asymptotic_var_alpha}. Secondly, we examine the differences between $I^{-1}JI^{-1}$ and $I^{-1}$. Since the asymptotic variance of a parameter in Bayesian inference follows $I^{-1}$ instead of $I^{-1}JI^{-1}$ \citep{frazier2023reliable}, discrepancies between them can result in overestimation or underestimation of the variance when they do not notice the model misspecification.

\subsection{Data Generation}

We generated the data $\bm{y}$ from the true model $p_n(\bm{y}|\bm{a}, \bm{b})$ in Eq.~\eqref{eqn:ncm_p_y}. We considered two cases: $K=2$ and $K=3$. In the case of $K=2$, we prepared $50$ problems, with $10$ problems for skill 1, $10$ problems for skill 2, and $30$ problems requiring both skill 1 and skill 2. For the case of $K=3$, we prepared $50$ problems, with $5$ problems for each skill, $5$ problems for each pair of skills, and the remaining $20$ problems requiring all three skills.

The item difficulty parameters for problems requiring only one skill were equally spaced from $-3$ to $3$ for $K=2$ and from $-2$ to $2$ for $K=3$. The item difficulty parameters for problems requiring two skills were drawn from a normal distribution with mean $-1$ and standard deviation $1.5$. The item difficulty parameters for problems requiring three skills were drawn from a normal distribution with mean $-1.5$ and standard deviation $1.5$. The item discrimination parameters were drawn from a log-normal distribution with parameters $\mbox{LogN}(0.2, 0.2)$.

To generate $\bm{y}$, we first drew $\bm{z}$ from the standard normal distribution. Then, we calculated the correct probability for each problem using Eq.~\eqref{eqn:ncm_p_y1}. Finally, we drew a response for each problem from a Bernoulli distribution, where the probability of a correct response was determined by the calculated correct probability.

\subsection{Simulation Settings}

To calculate $I^{-1}JI^{-1}$ and $I^{-1}$, we require the values of $\hat{\beta}_h^*$ and $\hat{\alpha}_{h,s}^*$. However, these values are not directly available. Therefore, we generated a large number of samples, specifically one million samples, and approximated the values using Eq.~\eqref{eqn:empirical_solution_in_cm}. Additionally, in order to calculate the expectations by $p_n(\bm{y})$ for $I^{-1}JI^{-1}$ and $I^{-1}$, we approximated them using 100k samples from $p_n(\bm{y})$.

For the calculation of the experimental asymptotic variances, we obtained $\hat{\beta}_h^{(n)}$ and $\hat{\alpha}_{h,s}^{(n)}$ by the samples from $p_n(\bm{y})$ with $n=10,000$. We repeated this process 30k times and calculated the experimental asymptotic variances.

To optimize the item parameters in Eq.~\eqref{eqn:empirical_solution_in_cm}, we utilized the {\tt mirt} library \citep{chalmers2012mirt}. In our preliminary experiments, we observed that the experimental asymptotic variances were consistently larger than the theoretical values. This phenomenon can be explained by two possibilities. The first possibility is that the marginal maximum likelihood estimation in Eq.~\eqref{eqn:empirical_solution_in_cm} constitutes a non-regular problem. If this is the case, Eq.~\eqref{eqn:asymptotic_var_beta} and Eq.~\eqref{eqn:asymptotic_var_alpha} do not hold. However, we also observed that the estimated parameters were distributed in a Gaussian shape, leading us to believe in the second possibility. The second possibility is that the optimization process terminated prematurely. Given the challenge of high-dimensional optimization such as 100 parameters, obtaining the exact optimal values is challenging. This introduces additional variance to the experimental asymptotic variance. To address this, we transformed the multivariate optimization into a one-parameter optimization approach. We initially performed the multivariate optimization to obtain $\hat{\beta}_h^{(n)}$ and $\hat{\alpha}_{h,s}^{(n)}$, and then conducted one-parameter optimization for each parameter. This involved initializing each parameter with a random value while keeping the remaining parameters fixed. This treatment yielded favorable results, as demonstrated in the subsequent section.

\subsection{Results}

\begin{figure}
 \centering
 {\tabcolsep=0pt
 \begin{tabular}{cc}
  \begin{minipage}{0.35\linewidth}
  \includegraphics[width=\linewidth]{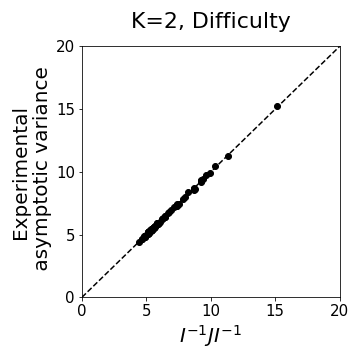}
  \end{minipage}
  &
  \begin{minipage}{0.35\linewidth}
  \includegraphics[width=\linewidth]{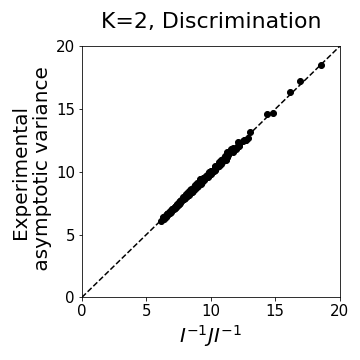}
  \end{minipage}\\
  \begin{minipage}{0.35\linewidth}
  \includegraphics[width=\linewidth]{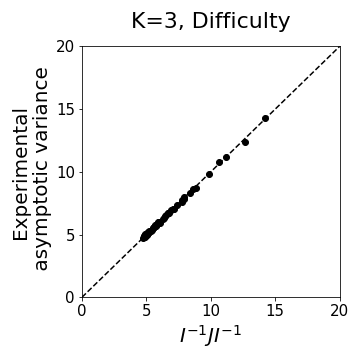}
  \end{minipage}
  &
  \begin{minipage}{0.35\linewidth}
  \includegraphics[width=\linewidth]{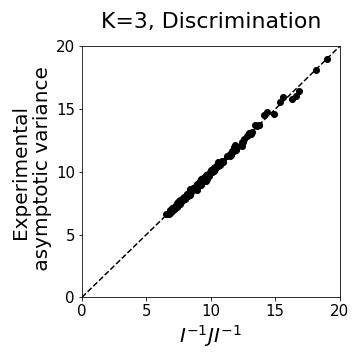}
  \end{minipage}
  \end{tabular}
  }
\caption{Scatter plots of $I^{-1}JI^{-1}$ and the experimental asymptotic variance for the item difficulty and discrimination parameters. The theoretical asymptotic variances $I^{-1}JI^{-1}$ are close to the experimental asymptotic variances.
}\label{pic:compare_infvar1_expvar}
\end{figure}

Firstly, we present the results comparing the experimental asymptotic variance and $I^{-1}JI^{-1}$ in Fig.~\ref{pic:compare_infvar1_expvar}. The scatter plots depict the item difficulty and discrimination parameters for both the $K=2$ and $K=3$ cases. We observe that $I^{-1}JI^{-1}$ and the experimental asymptotic variance are closely aligned in all cases. Table~\ref{tab:compare_infvar1_expvar} reports the mean absolute error (MAE) and mean absolute percentage error (MAPE) between $I^{-1}JI^{-1}$ and the experimental asymptotic variance. Both MAE and MAPE exhibit small values. The errors in the discrimination parameter tend to be larger than those in the difficulty parameter, and the $K=3$ case displays larger errors compared to the $K=2$ case. These results affirm the validity of Eq.~\eqref{eqn:asymptotic_var_beta} and Eq.~\eqref{eqn:asymptotic_var_alpha}.

\begin{figure}
 \centering
 {\tabcolsep=0pt
 \begin{tabular}{cc}
  \begin{minipage}{0.35\linewidth}
  \includegraphics[width=\linewidth]{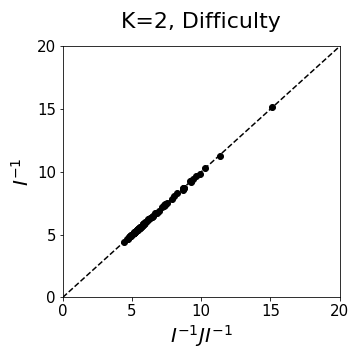}
  \end{minipage}
  &
  \begin{minipage}{0.35\linewidth}
  \includegraphics[width=\linewidth]{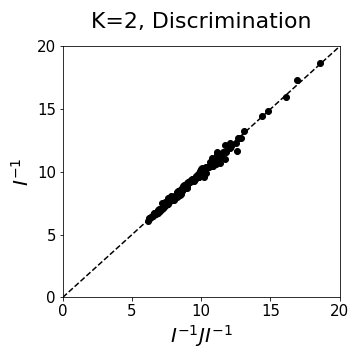}
  \end{minipage}\\
  \begin{minipage}{0.35\linewidth}
  \includegraphics[width=\linewidth]{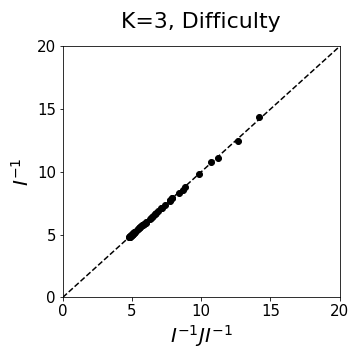}
  \end{minipage}
  &
  \begin{minipage}{0.35\linewidth}
  \includegraphics[width=\linewidth]{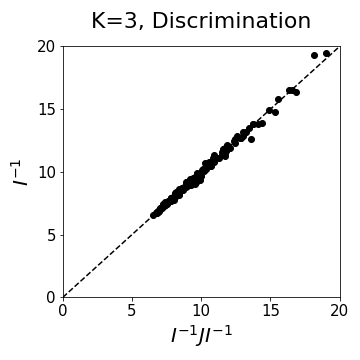}
  \end{minipage}
  \end{tabular}
  }
\caption{
Scatter plots of $I^{-1}JI^{-1}$ and $I^{-1}$ for the item difficulty and discrimination parameters. The theoretical asymptotic variances with and without considering misspecification are quite close.
}\label{pic:compare_infvar1_infvar2}
\end{figure}

Secondly, we present the results comparing $I^{-1}JI^{-1}$ and $I^{-1}$ in Fig.~\ref{pic:compare_infvar1_infvar2}. The scatter plots illustrate the item difficulty and discrimination parameters for both the $K=2$ and $K=3$ cases. We observe a close correspondence between $I^{-1}JI^{-1}$ and $I^{-1}$ in all cases. Table~\ref{tab:compare_infvar1_infvar2} reports the MAE and MAPE between $I^{-1}JI^{-1}$ and $I^{-1}$. Both metrics show small values. Similar to the previous comparison, the errors in the discrimination parameter are larger than those in the difficulty parameter. Moreover, the $K=3$ case exhibits larger errors compared to the $K=2$ case. These results indicate that $I^{-1}JI^{-1}$ and $I^{-1}$ are highly similar, suggesting neither overestimation nor underestimation occurs.

\begin{table}[tb]
    \centering
    \caption{Mean absolute error and mean absolute percentage error between $I^{-1}JI^{-1}$ and the experimental asymptotic variance for the item difficulty and discrimination parameters. Both metrics are quite small for all the cases.}
    \label{tab:compare_infvar1_expvar}
\begin{tabular}{lrrrr}
\toprule
& \multicolumn{2}{c}{$K=2$} & \multicolumn{2}{c}{$K=3$} \\
\cmidrule(r){2-3} \cmidrule(r){4-5}
& Difficulty & Discrimination & Difficulty & Discrimination \\
\midrule
MAE & 0.051 & 0.089 & 0.122 & 0.262 \\
MAPE (\%) & 0.721 & 0.911 & 0.832 & 1.153 \\
\bottomrule
\end{tabular}
\end{table}

\begin{table}[tb]
    \centering
    \caption{Mean absolute error and mean absolute percentage error between $I^{-1}JI^{-1}$ and $I^{-1}$ for the item difficulty and discrimination parameters. Both metrics are quite small for all the cases. }
    \label{tab:compare_infvar1_infvar2}
\begin{tabular}{lrrrr}
\toprule
 & \multicolumn{2}{c}{Dim=2} & \multicolumn{2}{c}{Dim=3} \\
 \cmidrule(r){2-3} \cmidrule(r){4-5}
 & Difficulty & Discrimination & Difficulty & Discrimination \\
\midrule
MAE & 0.023 & 0.128 & 0.049 & 0.282 \\
MAPE (\%) & 0.300 & 1.309 & 0.360 & 1.463 \\
\bottomrule
\end{tabular}
\end{table}

\section{Discussion}

Here, we discuss three advanced topics (higher dimensions, reverse misspecification, and correlated skills) related to the differences between estimated skills and true skills, as well as a practical implication of this paper.

First, this study focused on the two-dimensional case and clarified the mechanisms behind underestimation and overestimation. Eq.~\eqref{eqn:expected_gradient_learner_cm} still holds in higher dimensions; however, the case analysis we conducted in Fig.~\ref{pic:ncm_item_four_cases} becomes challenging due to the exponential growth of cases. Although a comprehensive understanding of higher dimensions is difficult, limited insights can be gleaned from the two-dimensional case analysis.
In Cases~1~and~2, the items exhibit imbalanced difficulty parameters, and we observed that the non-compensatory model classifies learners using all skills, while the compensatory model can be roughly viewed as classifying them using only more difficult skills (e.g., Skill~2 in Case~1). This approximation in the compensatory model is expected to occur in higher dimensions as well. Such items in higher dimensions have an underestimation effect on learners who possess high skills in the skills considered by the compensatory model and low skills in those not considered, leading to high $p_c$ and low $p_n$ ($p_n-p_c$ becomes negative). In Cases~3 and 4, where the items have balanced difficulty parameters, both models classify learners using all skills, and area C exists. In higher dimensions, one corner of a hypercube is truncated by the hyperplane of the compensatory model, leading to a weak overestimation effect on learners in area C. 
Although area C becomes extremely rare when each item is associated with many skills, such as 100, it is common for items to be associated with typically 1-3 skills, allowing area C to still exist.

Second, another area of interest is reverse misspecification, where the non-compensatory model is fitted to data from the compensatory model. Our approach to interpreting the gradient can also be applied in this reverse case by switching the compensatory model with the non-compensatory model. In this context, $p_n - p_c$ in Eq.~\eqref{eqn:expected_gradient_learner_cm} changes to $p_c - p_n$, and the four cases remain the same; thus, it is considered that underestimation and overestimation are flipped. It is anticipated that the lower skill is overestimated when one skill is high and the other is low. Additionally, two skills are expected to be slightly underestimated when both skills are at the same level and around zero. However, further analysis and experiments are required to conclude.

Third, considering situations where skills are correlated is practically important. It is known that when two skills are correlated, the compensatory and non-compensatory models yield more similar results \citep{christine2016partially}. This finding can be explained by the case analysis in Fig.~\ref{pic:ncm_item_four_cases}. As the correlation between skills increases, the number of learners in areas A and B decreases, leading to a reduction in underestimation. Area C also diminishes with increasing skill correlation, thereby decreasing overestimation. However, it is crucial to note that even in cases of skill correlation, if there is even one student with a high skill in one area and a low skill in another, the higher skill of that student will be underestimated. In critical assessment situations (e.g., entrance examinations), the impact on the student is significant.

Fourth, the practical contribution of this work lies in motivating model selection between compensatory and non-compensatory models. As pointed out by \cite{buchholz2018impact}, the compensatory model is commonly employed even when the skills assessed do not exhibit compensation. One possible reason for this is that the implications of model choice on estimated skills were previously unclear, leading users to overlook the significance of model selection and to default to the common compensatory model without careful consideration. Now that a comprehensive understanding of underestimation and overestimation has been provided, this result can be integrated into lectures and textbooks on MIRT models. We hope that our results and \cite{buchholz2018impact} will help users understand the importance of model choice and motivate them to select appropriate models for their specific situations.

\section{Conclusion}

We investigated the difference between the estimated skills and the true skills, as well as the variance of the estimated parameters when the non-compensatory model is mistakenly specified as the compensatory model. Concerning the difference, we obtained the gradient of the objective function of the compensatory model at the true skills, which allowed us to approximate the direction from the true skills to the estimated skills. By analyzing the gradient, we elucidated the mechanism underlying the occurrence of this difference. Specifically, the gradient provided insight into how the underestimation of higher skills arises, while we also discovered a slight overestimation phenomenon around the origin. Our experimental results confirmed a strong correlation between the gradient and the difference between the estimated skills and the true skills.

Regarding the variance, we derived the asymptotic variance of the parameters estimated by maximum likelihood estimation in our misspecified scenario. Through simulation studies, we validated that the derived asymptotic variance accurately assesses the actual variance. Moreover, we demonstrated that the derived asymptotic variance and the inverse of the Fisher information exhibit substantial similarity in typical datasets. Consequently, the underestimation or overestimation of the variance does not emerge as a critical concern in the assumed misspecified situation.


%
%
\section*{Data Availability}
Data used in this study is available from
\url{https://github.com/hrs-tamano/mirt_misspecification}.

\bibliographystyle{spbasic}      
\bibliography{main_with_authors}   


\end{document}